\documentclass[prb,reprint,aps,twocolumn]{revtex4-1}

\pdfoutput=1

\usepackage[utf8]{inputenc}

\usepackage[T1]{fontenc}
\usepackage{hyperref}
\usepackage{color}
\usepackage{graphicx}
\usepackage{units}
\usepackage{amsmath}

\usepackage{epstopdf}
\usepackage{subcaption}
\usepackage{amssymb}
\usepackage{pifont}
\usepackage{textcomp}
\usepackage[usenames,dvipsnames]{xcolor}
\usepackage{ulem}
\usepackage{amsmath} 

\usepackage{xfrac}

\hypersetup{colorlinks = true, citecolor = blue, breaklinks = true}

\begin{document}

\title{Bulk and surface properties of the Ruddlesden-Popper oxynitride Sr$_2$TaO$_3$N}
\date{\today}
\author{Maria Bouri}
\affiliation{Department of Chemistry and Biochemistry, University of Bern, Freiestrasse 3, CH-3012 Bern, Switzerland}
\author{Ulrich Aschauer}
\affiliation{Department of Chemistry and Biochemistry, University of Bern, Freiestrasse 3, CH-3012 Bern, Switzerland}

\begin{abstract}
Oxynitrides with the perovskite structure are promising candidates for photocatalysis under visible light due to their appropriate optical and electronic properties. Recently, layered perovskites have attracted attention for their improved performance with respect to the bulk perovskites in photocatalytic water splitting. In this paper, we investigate the structural and electronic properties of the layered Ruddlesden-Popper oxynitride Sr$_2$TaO$_3$N and its (001) surfaces using density functional theory (DFT) calculations. We find that the energetically favoured configuration of the bulk has an in-plane \textit{cis} anion order and exhibits rotations of the TaO$_6$ octahedra. Furthermore, we show that the TaON-terminated (001) surface suppresses exciton recombination due to higher-energy surface states, giving a potential explanation for the good photocatalytic performance.
\end{abstract}

\maketitle
\section*{Introduction}

Perovskite oxides represent an important class of materials due to their wide range of applications such as ferroelectric and magnetic functional materials and solid electrolytes. Perovskite oxides can also serve as photocatalysts for water splitting\cite{Wang2015,Kato2003} to produce hydrogen as a clean and renewable energy carrier\cite{Lewis2001}. Their wide band gap, however, restricts light absorption to the ultraviolet part of the solar spectrum. One route to reduce the band gap and achieve absorption of visible light - required for economically viable hydrogen production - is doping/alloying perovskite oxides with foreign elements. The introduction of foreign cations has been widely studied\cite{Kudo2009}, while substitution on the anion site has only been investigated more recently. Among these anion substituted materials, oxynitrides are derived from pure oxides by a partial substitution of oxygen by nitrogen. Nitrogen is less electronegative than oxygen and introduces 2\textit{p} orbitals at higher energies, thus raising the top of the valence band, narrowing the band gap\cite{Castelli2013a} and leading to visible-light absorption \cite{Ji2005a}. Nitrogen is also more polarisable than oxygen and forms stronger covalent bonds with the metal, which lowers the metal \textit{d}-derived conduction band states\cite{Fuertes2012}, further reducing the band gap and making perovskite oxynitrides promising photocatalysts\cite{Takata2015,Umezawa2016,Pan2015}. 

Layered structures are a further modification of perovskite materials that can lead to distinctively different properties compared to the bulk material. Compared to chemically similar non-layered bulk perovskites, the photocatalytic water-splitting activity seems to be improved for these layered materials\cite{Kudo2009}. Among these layered structures, Ruddlesden-Popper phases with the general formula A$_{n+1}$B$_n$O$_3$ are composed of $n$ perovskite blocks separated by intertwining rock-salt layers\cite{Ruddlesden1957,Beznosikov2000}.

\begin{figure}
	\centering
	\includegraphics[width=0.6\columnwidth]{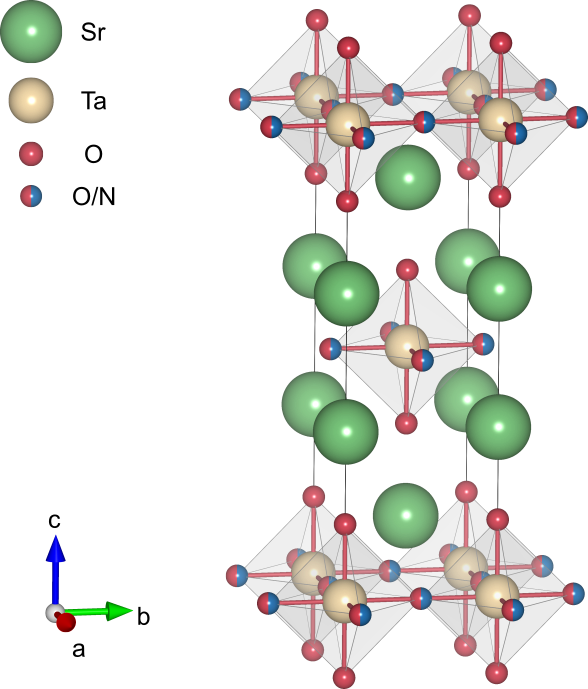}
	\caption{14-atom unit cell of the high-symmetry Sr$_2$TaO$_3$N structure.}
	\label{fig:bulk_14_atom}
\end{figure}

In this study we aim to investigate the simultaneous effect of nitrogen substitution and a layered Ruddlesden-Popper structure on the electronic properties by means of density functional theory (DFT) calculations. Although the crystal structures of some Ruddlesden-Popper oxynitrides have been experimentally resolved \cite{Clarke2002,Tobias2001,Tobias2004a} and a computational high-throughput screening study confirmed the suitability of their band gaps and band edges for photocatalytic water splitting\cite{Castelli2013a}, details about their electronic structure, in particular at surfaces, are still unknown. Here we investigate Sr$_2$TaO$_3$N (Fig. \ref{fig:bulk_14_atom}) as an experimentally synthesised \cite{Diot1999,Clarke2002} prototype material of this class, along with its (001) surface. After establishing the energetically favoured nitrogen anion order and crystal structure of bulk Sr$_2$TaO$_3$N, we investigate the electronic structure of the bulk and symmetric SrO- and TaON-terminated (001) surfaces. Our results show that the TaON-terminated slab may prevent electron-hole recombination as high-lying surface states preferentially accommodate holes, while electrons reside in the subsurface and bulk layers. This natural charge separation may provide an explanation for the good performance of Ruddlesden-Popper oxynitride photocatalysts.

\section*{Methods}
Our plane-wave density functional theory (DFT) calculations were performed with the Quantum ESPRESSO package\cite{Giannozzi2009} using the generalised gradient approximation (GGA) within the Perdew-Burke-Ernzerhof (PBE)\cite{Perdew1996} exchange correlation functional. Electronic wave functions were expanded in a plane-waves  with a kinetic-energy cutoff of 35 Ry combined with 280 Ry for the augmented density. Ultrasoft pseudopotentials\cite{Vanderbilt1990} were used with Sr(4s, 4p, 4d, 5s, 5p), Ta(5s, 5p, 5d, 6s, 6p), O(2s, 2p), and N(2s, 2p) valence states.

The initial high-symmetry Sr$_2$TaO$_3$N Ruddlesden-Popper structure, visualised in Fig. \ref{fig:bulk_14_atom} using VESTA\cite{Momma2011}, was taken from Diot \textit{et al.}\cite{Diot1999} where nitrogen ions occupy half of the equatorial sites, while the apical sites are occupied only by oxygen. Our calculations confirmed this preference of N for the equatorial sites and we thus investigated all possible explicit nitrogen arrangements on the equatorial sites within a $2\times2\times1$ supercell.

The structures were optimised until forces converged below 0.025 eV/\AA. The Brillouin zone was sampled using $8\times8\times2$ and $4\times4\times2$ Monkhorst-Pack\cite{Pack1977} \textit{k}-points meshes for the 14-atom unit cell and its $2\times2\times1$ supercell (56 atoms) respectively. A denser \textit{k}-point mesh of $8\times8\times4$ was used for the projected density of states (PDOS) calculations of supercells within PBE. For the determination of band gaps, we used the screened hybrid functional HSE06\cite{Heyd2006} on relaxed PBE structures using - for computational feasibility - a uniaxially less dense $8\times8\times2$ \textit{k}-mesh.  

Phonons were computed using the frozen-phonon method\cite{Kunc1982} within the PHONOPY code\citep{Togo2015}. To determine dynamical instabilities also in the perfectly disordered system, phonon calculations were also carried out for a structure with a 50/50 O/N virtual-crystal approximation (VCA) \cite{Bellaiche2000} pseudopotential for the equatorial anion sites. Effective masses were calculated with the finite difference method implemented in the Effective Mass Calculator (EMC) program\cite{Fonari}.

Surfaces along the (001) direction were constructed from the optimised 56-atom bulk structure as slabs using a vacuum of 10 \AA\ to separate periodic images along the surface-normal coordinate. The SrO termination is obtained by cleaving between adjacent SrO layers and results in a symmetric and stoichiometric 22.6 \AA\ thick slab with 12 atomic layers. In this case the middle two SrO layers were kept fixed at bulk positions during relaxation. The TaON termination is modelled in a symmetric but nonstoichiometric 13 atomic layer thick slab (24.8 \AA) with one additional TaON layer. Here the middle 3 atomic layers were fixed at bulk positions, resulting in the same number of free layers for the two terminations. A $4\times4\times1$  \textit{k}-point mesh was used for Sr$_2$TaO$_3$N (001) surface calculations.

\section*{Results and discussion}

\subsection*{Bulk properties}
Many perovskites and layered perovskites show deviations from the perfect cubic structure, most commonly antiferrodistortive (octahedral rotations) or ferroelectric (polar) distortions \cite{Benedek2013,Birenbaum2014}. We thus investigate structural instabilities in high-symmetry Sr$_2$TaO$_3$N with the frozen phonon method, for the structure with anion disorder on the equatorial sites (modelled via the VCA) as well as for \textit{trans} and \textit{cis} nitrogen anion orders on the equatorial sites. Table \ref{tbl:table1} shows the frequencies of the unstable phonons as well as the energy gain associated with relaxing the instability. For non-VCA calculations we also show the relative energies of the relaxed $2\times2\times1$ supercells with different anion orders.

\begin{table}
\small
	\caption{Frequencies of unstable phonons, their associated energy lowering and the energy of the relaxed $2\times2\times1$ supercell relative to the most favourable structure.}
	\label{tbl:table1}
	\begin{tabular*}{0.5\textwidth}{cc|ccc}
	\hline
	Anion & Instability & Frequency & Energy gain & Relative energy\\
	order & & (THz) & (eV) & (eV)\\
	\hline
	VCA & Rotation & 2.08\textit{i} & 0.060 &\\
	VCA & Polar & 4.65\textit{i} & 0.001 &\\\hline
	\textit{trans} & Rotation & 2.51\textit{i} & 0.155 & 2.32\\
	\textit{trans} & Polar & 4.80\textit{i} & 0.006 & 2.47\\
	\textit{cis} & Rotation & 2.15\textit{i} & 0.120 & 0.00\\
	\end{tabular*}
\end{table}

For O/N disorder on the equatorial sites modelled via the VCA we obtain imaginary modes both at the Brillouin zone boundary and at the $\Gamma$ point, corresponding to octahedral rotations and an in-plane polar distortion respectively. The energy gain associated with these instabilities is small, the rotational instability, however, is significantly more favourable than the polar one.

\begin{figure}
	\centering
	\includegraphics[width=0.8\columnwidth]{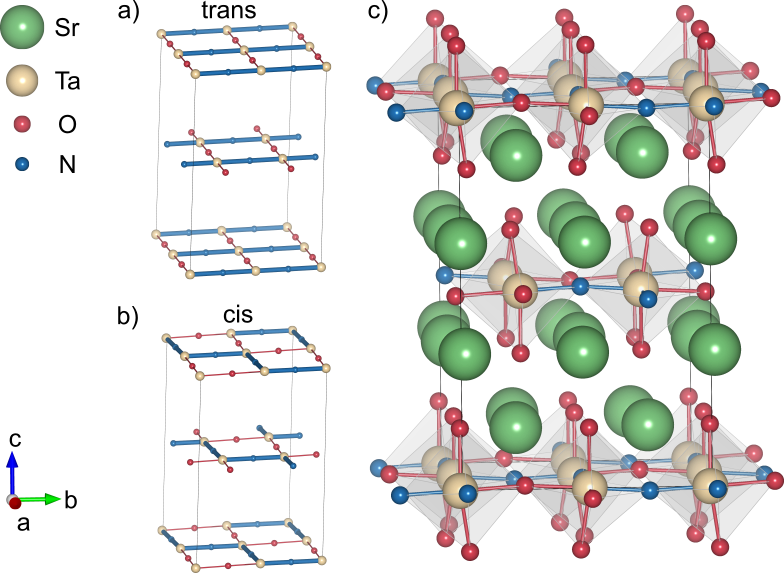}
	\caption{\textit{Left}: Schematic representation of the most stable Sr$_2$TaO$_3$N 56-atom undistorted unit cell for a) \textit{trans} and b) \textit{cis} nitrogen orders and c) the energetically most favourable Sr$_2$TaO$_3$N 56-atom unit cell with \textit{cis}-anion order and octahedral rotations.}
	\label{fig:bulk_56_atom}
\end{figure}

Due to the layered structure of Sr$_2$TaO$_3$N, different orientations of \textit{cis} (N-Ta-N bonds form 90 angles) and \textit{trans} (N-Ta-N bonds form 180 angles) nitrogen order per layer are possible. Therefore, we consider all possible combinations of parallel or perpendicular N-Ta-N chains within adjacent perovskite layers of the Sr$_2$TaO$_3$N structure. After relaxation, the energetically most favourable \textit{cis} and \textit{trans} arrangements consist of parallel N-Ta-N chains in the two layers as shown in Figs. \ref{fig:bulk_56_atom}a) and b). For the energetically most favourable \textit{trans} N atom arrangement we find unstable octahedral rotation and polar modes. The energy gain associated with the rotational instability is again larger than the one of the polar instability, as we already observed for the disordered structure. Arranging the N atoms in a \textit{cis} order we find a single imaginary phonon mode representing a rotational instability, the polar instability that existed with VCA and \textit{trans} order being suppressed in the \textit{cis} arrangement.

Overall, the relaxed energy of bulk Sr$_2$TaO$_3$N with \textit{cis} order and rotational distortions is significantly lower than the energy of the \textit{trans}-ordered structures (Table \ref{tbl:table1}). This is in a good agreement with several studies on non-layered tantalum oxynitrides with perovskite structure that prefer a \textit{cis} over a \textit{trans} nitrogen arrangement due to the larger hybridisation of nitrogen \textit{p} states  with tantalum \textit{d} states, which leads to stronger covalent bonds\cite{Yang2011}.

\begin{figure}
	\centering
	\includegraphics[width=\columnwidth]{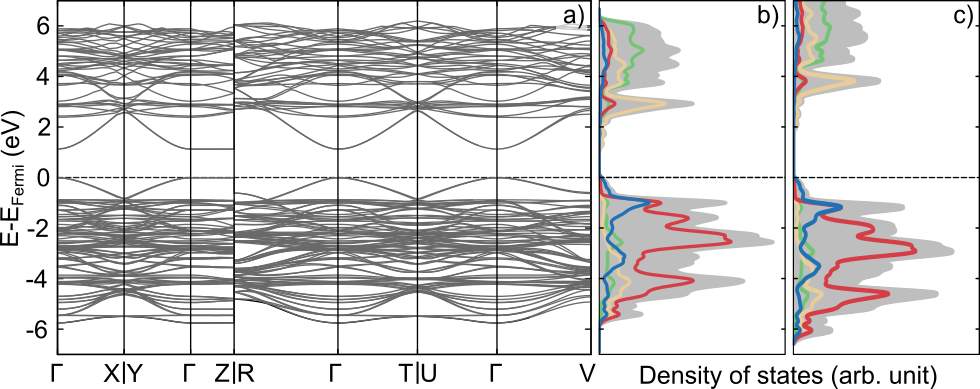}
	\caption{a) Electronic band structure using PBE and b) corresponding electronic DOS of the energetically favoured configuration of bulk Sr$_2$TaO$_3$N. Panel c) shows the electronic DOS using HSE06. The Fermi energy is set to the top of the valence band (dashed line). The high-symmetry points in a) are: $\Gamma$ = (0 0 0), X = ($\sfrac{1}{2}$ 0 0), Y = (0 $\sfrac{1}{2}$ 0), Z = (0 0 $\sfrac{1}{2})$, R = (-$\sfrac{1}{2}$ -$\sfrac{1}{2}$ $\sfrac{1}{2}$), T = (0 -$\sfrac{1}{2}$ $\sfrac{1}{2}$), U = -($\sfrac{1}{2}$ 0 $\sfrac{1}{2}$), V = ($\sfrac{1}{2}$ -$\sfrac{1}{2}$ 0).}
	\label{fig:dos_bands}
\end{figure}

In Figure \ref{fig:dos_bands}, we show the electronic band structure and density of states (DOS) of the most stable \textit{cis}-ordered Sr$_2$TaO$_3$N bulk configuration. The projected DOS (Fig.\ref{fig:dos_bands}b) reveals that the top of the valence band mainly consists of nitrogen states while oxygen states are located at lower energies. This observation is in agreement with other theoretical studies on oxynitrides as nitrogen, due to its lower electronegativity, results in higher energy \textit{p} states that raise the top of the valence band and result in a narrower band gap compared to pure oxides\cite{Ji2005a}. Moreover, the bottom of the conduction band is mainly formed by tantalum 5\textit{d} states while the strontium states are located at higher energies. 

The electronic band structure of Sr$_2$TaO$_3$N (Fig. \ref{fig:dos_bands}a) shows strongly dispersive bands in the \textit{xy} plane and flat bands along the $\Gamma$ - Z direction in both valence and conduction band, which hints towards large effective masses for the hole and the electron along this direction. We calculate effective masses for holes along the \textit{x} and \textit{y} axis of -0.205$m_{0}$ and -0.767$m_{0}$ respectively ($m_0$ being the rest mass of the electron) indicating good mobility of the hole in the \textit{xy} plane. Light effective masses are also computed for electrons (0.272$m_{0}$ and 0.228$m_{0}$ along the \textit{x} and \textit{y} axis respectively). However, the effective masses along the \textit{z} direction for both electrons and holes are large, approaching infinity, implying a low mobility along this direction. The layered structure leads to a weakened bonding of Ta t$_{2g}$ orbitals with a z component (i.e. $d_{xz}$ and $d_{yz}$), which become less dispersive and appear as the marked peak above 2eV in the DOS. Due to the complete in-plane N order, the $d_{xy}$ band should be more dispersive and extend to lower energies than in a comparable bulk perovskite (i.e. SrTaO$_2$N).

It is known that semi-local DFT functionals underestimate the band gap and we thus performed calculation with the HSE06 hybrid functional to support the PBE results. The resulting DOS (Fig. \ref{fig:dos_bands}c) is similar to the one obtained using PBE, the primary change being a widening of the band gap (E$_{bandgap}^{PBE}$ = 1.128 eV, E$_{bandgap}^{HSE}$ = 2.005 eV) as unoccupied conduction band states shift to higher energies. Therefore, besides predicting too small a band gap, our PBE results should correctly describe the electronic band structure. The HSE calculations confirm that Sr$_2$TaO$_3$N has a suitable band gap for photocatalytic water splitting, which is larger than the minimal requirement of 1.23 eV but still small enough to absorb a significant fraction of visible light.

\subsection*{Surface properties}

The electronic properties of the surfaces of Sr$_2$TaO$_3$N are important in order to understand the suitability of the material as a photoelectrode. During photocatalysis, the adsorbed photon will excite an electron to the conduction band, leaving a hole in the valence band. The photogenerated electrons and holes will migrate to the surface to contribute to the reduction and oxidation respectively of the adsorbed species, ultimately resulting in the production of H$_2$ and O$_2$. A good photocatalyst will prevent the recombination of the electron with the hole in the bulk and at the surface. Surface states have the potential to suppress electron-hole recombination if they lead to a spatial separation of the carriers.

\begin{figure}
	\centering
	\includegraphics[width=\columnwidth]{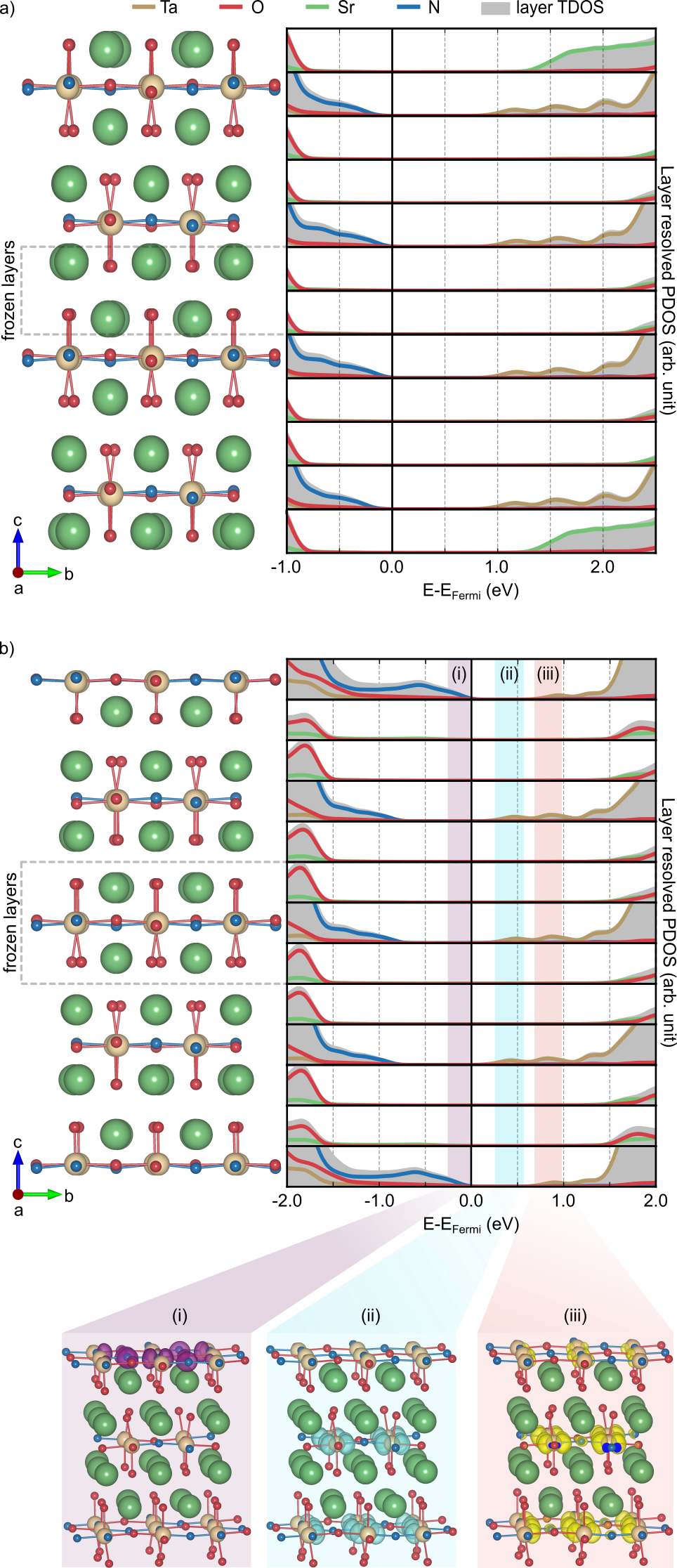}
	\caption{Layer resolved electronic density of states for a) SrO- and b) TaON-terminated surfaces. The two middle SrO layers and the three middle layers respectively were kept fixed at bulk positions. The panels i)-iii) show local density of states integrated in the energy ranges highlighted in panel b).}
\label{fig:edos_surface}
\end{figure} 

We investigate the electronic structure of the (001) surfaces with both SrO and TaON terminations (Fig. \ref{fig:edos_surface}). The densities of states of the bulk layers for both the SrO- and TaON-terminated slabs are virtually identical to the bulk case. In the case of the SrO-terminated slab (Fig. \ref{fig:edos_surface}a), we observe primarily changes in the very surface layer, where conduction band Sr states shift to lower energies compared to the bulk layers. These states, however, remain above the tantalum states of the subsurface and bulk TaON layers, indicating that electrons will preferentially fill these unoccupied Ta states. Given that there is no major shift of the top of the valence band as a function of the layer depth, holes will occupy subsurface and bulk states with the same likelihood. This implies that electrons and holes occupy the same spatial regions, without any suppression of recombination compared to the bulk.

In the TaON-terminated slab, the surface valence band is shifted to higher energies, indicating that holes will preferentially be located in the surface layer (Fig. \ref{fig:edos_surface}b). The integrated local density of states (ILDOS) of the top of the valence band confirms that only the nitrogen states of the surface layer contribute in this range (Fig. \ref{fig:edos_surface}b (i)). On the other hand, the strongly dispersive band forming the conduction band edge in the bulk is less dispersive at the surface, which leads to higher-energy empty states in the surface region and a preferential accommodation of excess electrons in bulk and subsurface layers. The ILDOS for the energy ranges of the two states in the conduction band clearly reveals that the lower energy range is composed of the bulk tantalum \textit{d} states (Fig. \ref{fig:edos_surface}b (ii)) while the surface tantalum states are located at higher energies (Fig. \ref{fig:edos_surface}b (iii)). This preferential accommodation of holes at the surface and electrons in the subsurface and bulk implies that surface recombination is suppressed by the nature of the surface states on the TaON termination.

Despite this electronic structure advantage, the potential performance of Sr$_2$TaO$_3$N (001) surfaces as a photoelectrodes will likely be hindered by the low carrier mobility along the bulk to surface direction. It is however possible that polycrystalline materials, in which the low conductivity direction changes from grain to grain could significantly enhance the bulk to surface charge transport while still excluding electrons from the surface layers. We would therefore expect a poorer performance of single-crystalline compared to polycrystalline Sr$_2$TaO$_3$N photoelectrodes. Alternatively, the higher carrier mobility could render smaller facets perpendicular to the [001] direction more reactive, which is ongoing work.

\section*{Conclusions}
We investigated the structural and electronic properties of bulk Sr$_2$TaO$_3$N and its (001) surfaces using DFT. We show that octahedral rotation instabilities are favoured over polar distortion independently of the anion order. The energetically most favourable anion order in the bulk is a \textit{cis} arrangement with N-Ti-N bonds forming a \textit{zig-zag} pattern with parallel orientation in the two layers. The projected DOS show that nitrogen states form the top of the valence band, narrowing the band gap compared to pure oxides, which is in agreement with previous studies on oxynitrides. The bottom of the conduction band is mainly formed by tantalum states with strontium states located at higher energies.

We obtain similar behaviour in the projected DOS of the bulk layers of the SrO- and TaON- terminated (001) surface slabs while surface states are observed in their top-most layers. In the case of the SrO-terminated slab holes and excess electrons will preferentially be accommodated in bulk layers without any suppression of recombination. On the other hand for the TaON-terminated slab, we find that surface-layer nitrogen states at the valence-band edge shift to higher energies implying the surface's preference to accommodate holes. Simultaneously, the tantalum states forming the conduction-band edge of the surface layer are higher in energy, which excludes excess electrons from the surface layer. This result reveals that TaON-terminated slab suppress surface electron-hole recombination, making it a promising surface for photocatalysis. Due to the low carrier mobility along the surface normal direction, we predict polycrystalline materials to have a higher photocatalytic activity than single crystals.

\section*{Acknowledgements}
This research was funded by the SNF Professorship Grant PP00P2\_157615. Calculations were performed on UBELIX (http://www.id.unibe.ch/hpc), the HPC cluster at the University of Bern.

\bibliography{library.bib}

\end{document}